\documentclass[journal=jacsat,manuscript=article]{achemso}

\usepackage{amssymb}
\usepackage{amsmath}
\usepackage[labelsep=period]{caption}
\usepackage{hyperref}
\hypersetup{
    colorlinks,%
    citecolor=blue,%
    linkcolor=blue,%
    urlcolor=blue
}

\usepackage{chemformula} 
\usepackage[T1]{fontenc} 

\author{Ruiqi Hu}
\affiliation{Department of Materials Science and Engineering, University of Delaware, Newark, DE 19716, USA}
\email{ruiqihu@udel.edu}

\author{Dai Q. Ho}
\affiliation{Department of Materials Science and Engineering, University of Delaware, Newark, DE 19716, USA}
\alsoaffiliation{Faculty of Natural Sciences, Quy Nhon University, Quy Nhon 590000, Vietnam}

\author{Quang To}
\affiliation{Department of Materials Science and Engineering, University of Delaware, Newark, DE 19716, USA}

\author{Garnett W. Bryant}
\affiliation{Nanoscale Device Characterization Division, Joint Quantum Institute, National Institute of Standards and Technology, Gaithersburg, MD 20899-8423, USA and the University of Maryland, College Park, MD 20742, USA}
\email{garnett.bryant@nist.gov }

\author{Anderson Janotti}
\affiliation{Department of Materials Science and Engineering, University of Delaware, Newark, DE 19716, USA}
\email{janotti@udel.edu}

\title[An \textsf{achemso}]
  {Fermi-level pinning in ErAs nanoparticles embedded in III-V semiconductors}

\begin{document}

\begin{abstract}

Embedding rare-earth pnictide (RE-V) nanoparticles into III-V semiconductors enables unique optical, electrical, and thermal properties, with applications in THz photoconductive switches, tunnel junctions, and thermoelectric devices. Despite the high structural quality and control over growth, particle size, and density, the underlying electronic structure of these nanocomposite materials has only been hypothesized. Basic questions about the metallic or semiconducting nature of the nanoparticles (that are typically < 3 nm in diameter) have remained unanswered. Using first-principles calculations, we investigated the structural and electronic properties of ErAs nanoparticles in AlAs, GaAs, InAs, and their alloys. Formation energies of the ErAs nanoparticles with different shapes and sizes (i.e., from cubic to spherical, with 1.14 nm, 1.71 nm, and 2.28 nm diameters) show that spherical nanoparticles are the most energetically favorable. As the diameter increases, the Fermi level is lowered from near the conduction band to the middle of the gap. For the lowest energy nanoparticles, the Fermi level is pinned near the mid-gap, at about 0.8 eV above the valence band in GaAs and about 1.2 eV in AlAs, and it is resonant in the conduction band in InAs. Our results show that the Fermi level is pinned on an absolute energy scale once the band alignment at AlAs/GaAs/InAs interfaces is considered, offering insights into the rational design of these nanocomposite materials.

\end{abstract}

\section{Introduction}

\begin{figure}[htp]
\centering
\includegraphics[width=\textwidth]{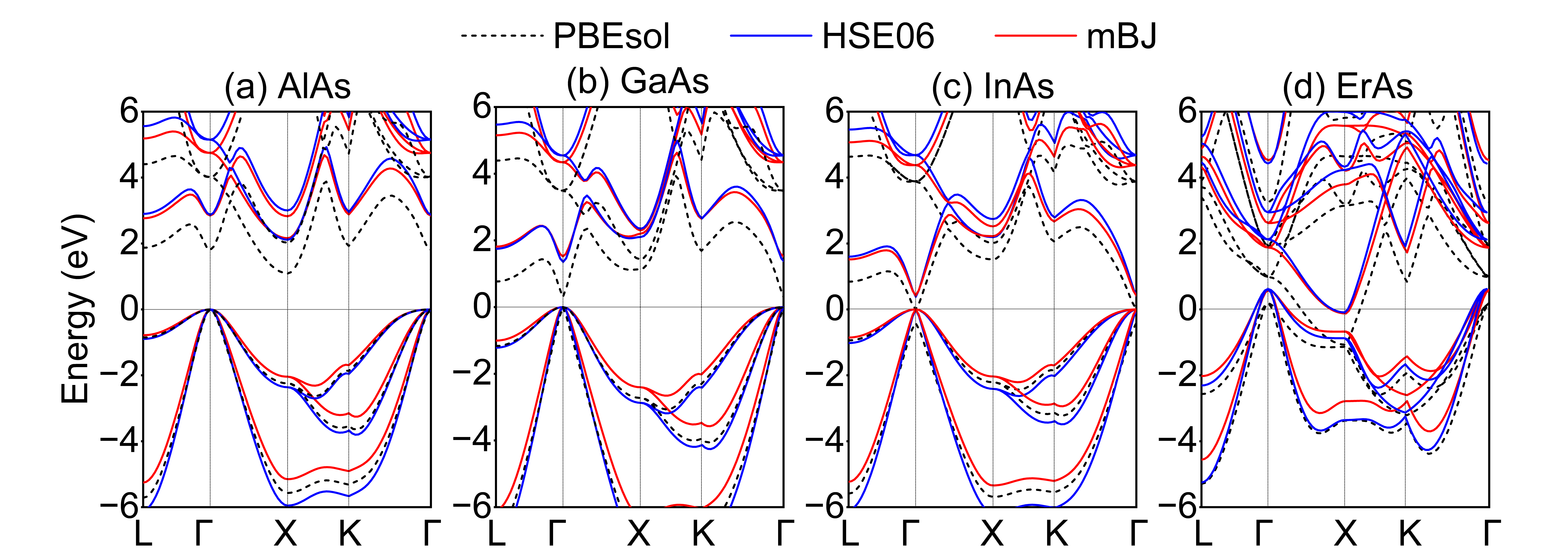}
\captionsetup{width=\textwidth}
\caption{\label{fig1}
Electronic band structures of (a) AlAs, (b) GaAs, (c) InAs, and (d) ErAs with three different functionals, PBEsol, HSE06 and meta-GGA mBJ. The zero on the energy axis is set to the valence-band maximum $E_v$ in the semiconductors and the Fermi level in ErAs (i.e., not aligned according to the band offsets).}
\end{figure}

Rare-earth monopnictides (RE-V, where RE=La,..., Lu, and V=As, Sb, Bi) have gained significant attention in recent years due to their potential applications in THz photoconductive switches\cite{taylor2006resonant}, thermoelectrics \cite{kim2006thermal}, tunnel junctions\cite{nair2010enhanced}, and epitaxially perfect metallic contacts\cite{zimmerman2005tunable}. With a simple rocksalt crystal structure, RE-Vs display a characteristic overlap of the RE-derived 5$d$ bands (electron pockets) and the pnictide $p$ bands (hole pockets) as depicted in Fig.~\ref{fig1}(d), making them semimetallic. RE-Vs have been grown as bulk single crystals, thin films, and embedded nanoparticles within conventional III-V semiconductors \cite{kawasaki2013surface,bomberger2017overview}. Notably, epitaxial thin films of RE-Vs such as ErAs, TbAs, GdSb, and LuSb on substrates like GaAs, InGaAs, InAs, and GaSb have been successfully grown using molecular beam epitaxy (MBE)\cite{palmstrom1988epitaxial,allen1989eras,klenov2005interface,cassels2011growth,clinger2012thermoelectric,chatterjee2019weak,inbar2022epitaxial}. Coherent thin films with high structural quality were grown down to a few monolayers by appropriately matching the rocksalt RE-V to the zinc blende III-V lattice parameters. Incorporating semi-metallic RE-V nanoparticles in III-V semiconductor matrices has demonstrated the capability to modify their optical, electrical, and thermoelectric properties, with significant implications for device engineering \cite{lewis2018size}. Remarkably, the growth of RE-V nanoparticles within III-V host matrices utilizing MBE has enabled unprecedented control over their positions, sizes, and density \cite{lewis2018size}. However, the underlying electronic structure of these nanocomposite materials has remained a subject of debate\cite{lewis2018size}, with measurements of structural, thermal, electrical, optical, and magnetic properties providing only indirect evidence. 

The incorporation of nanoparticles into conventional III-V semiconductors leads to significant modifications, including reduced carrier lifetime \cite{kadow1999}, elevated phonon scattering \cite{zide2005thermoelectric}, decreased thermal conductivity, yet enhanced electrical conduction through electron filtering mechanisms \cite{nair2010enhanced}. Despite the different crystal structures of the RE-Vs and the III-Vs, (rocksalt vs.~zinc blende), their crystalline lattices nicely match by having a common anion (group-V) sublattice, differing only in the position of the metal atom. For example, there exists a small lattice mismatch of 1.6\% between ErAs and GaAs \cite{young2014lattice}; observations from both experiments and density functional theory calculations of GaAs/ErAs(001) interfaces have established that the Ga-terminated interface, known as the chain model, with As forming a continuing sublattice across the (001) interface is the more energetically favorable interface termination \cite{lambrecht1998schottky,klenov2005interface,delaney2010theoretical}.
Measurements of cross-sectional scanning tunneling microscopy (STM) on ErAs nanoparticles embedded in GaAs (ErAs:GaAs) show high structural quality and confirm the existence of a continuous As sublattice \cite{kawasaki2011local}, while results of scanning tunneling spectroscopy (STS) indicate a Fermi level pinning near midgap of the otherwise heavily n-type doped GaAs (5$\times$10$^{18}$ cm$^{-3}$ Si) \cite{kawasaki2011local}. However, these observations lack theoretical justification. 

Employing first-principles calculations based on density functional theory, we explore the structural and electronic properties of ErAs nanoparticles embedded in GaAs, AlAs, InAs, and their alloys. We consider different nanoparticle shapes (from cubic to spherical) and sizes (from 1.14 nm to 2.28 nm in diameter). We find that spherical nanoparticles are energetically preferred, and the Fermi level is pinned at higher energies in the gap for the smaller nanoparticles. The Fermi level is pinned at lower energies as the nanoparticle size increases. Remarkably, for nanoparticles with the same size but in different III-V hosts, the Fermi level is pinned on an absolute energy scale (i.e., with respect to the vacuum level) when the natural band offsets between the different III-Vs are taken into account.

\section{Computational methods}

The calculations are based on the density functional theory (DFT) \cite{Hohenberg1964,KohnSham1965} as implemented in the VASP code \cite{kresse1993ab, kresse1994ab}. The interactions between the valence electrons and the ions are treated using the projector augmented wave (PAW) method \cite{blochl1994projector,kresse1999ultrasoft}. The Er 4f orbitals were treated as core orbitals, as the 4f occupied and unoccupied bands in ErAs are sufficiently far from the Fermi level \cite{Shoaib2020}. The equilibrium lattice parameters of the III-V host materials and the RE-V were calculated using the generalized gradient approximation for the exchange and correlation as in the revised Perdew-Burke-Ernzerhof functional PBEsol \cite{perdew2008restoring} for the two-atoms primitive cell and an 8$\times$8$\times$8 mesh of $k$ points. All calculations were performed with a 300 eV cutoff energy for the plane-wave basis set.

The electronic band structures were calculated using the PBEsol, HSE06 hybrid functional\cite{hse06-2003,hse06}, and the modified Becke-Johnson meta-GGA (mBJ) functional \cite{becke2006simple,tran2009accurate}. The results are shown in Fig.~\ref{fig1}. The PBEsol functional not only underestimates the band gap of semiconductors, it also overestimates the overlap between electron and hole pockets in the semimetal RE-V compounds \cite{Shoaib2020}. The HSE06 hybrid functional gives band gaps that are in much better agreement with experimental values, and corrects the overlap between the electron and hole pockets in the RE-V \cite{heyd2005,Shoaib2020}. The results of the mBJ functional are very close to that of the HSE06 functional, attesting to its accuracy, yet at a much lower computational cost. Note that the atom-dependent parameter $c_{\text{mBJ}}$ in the mBJ functional, was slightly modified to improve the description of the the overall band structures of the various materials considered here, with values of 1.176, 1.267, 1.220, 1.160, and 1.143 for Al, Ga, In, As, and Er. Previous calculations indicated that the band gap of semiconductors increases monotonically with respect to $c_{\text{mBJ}}$ \cite{tran2009accurate}.
It should also be noted that the mBJ functional cannot achieve self-consistency with respect to the total energy and, therefore, cannot be used to compute the Hellmann-Feynman forces, (i.e., relax the atomic positions).

The embedded ErAs nanoparticles were simulated using a 7$\times$7$\times$7 repetition of the host III-V (AlAs, GaAs, and InAs) 8-atom conventional unit cell resulting in a total 2744 atoms for the III-V host supercell. The positions of all the atoms in the supercell were allowed to relax, while maintaining the lattice parameters of the parent III-V materials. These calculations were performed using only $\Gamma$ for sampling the Brillouin zone and the PBEsol functional. The calculations for the density of states of the composite material, used for determining the Fermi level position, were performed using the mBJ functional with a $\Gamma$-centered 4$\times$4$\times$4 mesh of $k$ points.

\section{Results and discussion}

The calculated equilibrium lattice parameters using the DFT-PBEsol functional are 5.681 \AA~for AlAs, 5.683 \AA~for GaAs, 6.123 \AA~for InAs, and 5.688 \AA~for ErAs, in good agreement with experimental data (5.661 \AA, 5.653 \AA, 6.058 \AA, and 5.743 \AA, respectively \cite{madelung2004semiconductors,palmstro1988epitaxial}). 
For comparison, the equilibrium lattice parameters using the HSE06 functional are 5.676 \AA~for AlAs, 5.667 \AA~for GaAs, 6.108 \AA~for InAs, and 5.737 \AA~for ErAs.

The computed band structures using PBEsol, HSE06, and meta-GGA mBJ functionals are presented in Fig.~\ref{fig1}. As expected, the calculated band gaps for AlAs, GaAs, and InAs using PBEsol are severely underestimated; the HSE06 functional gives band gaps in good agreement with the experimental data\cite{vurgaftman2001band,madelung2004semiconductors}, and the meta-GGA mBJ provides band structures and band gaps that are close to those in HSE06. For the semimetal ErAs, the PBEsol overestimates the overlap in energy between the electron pockets at the X point and hole pockets at the $\Gamma$ point \cite{Shoaib2020}, leading to an artificially higher carrier concentration compared to experimental data\cite{bogaerts1996experimental}. 
In contrast, both HSE06 and mBJ functionals exhibit an energy overlap of approximately 0.71 eV, and give carrier concentrations in much better agreement with experiment \cite{bogaerts1996experimental}. These results give us confidence in the results for the electronic structure calculations of the nanoparticle/III-V composite materials using mBJ.

\begin{figure}
    \centering
    \includegraphics[width=\textwidth]{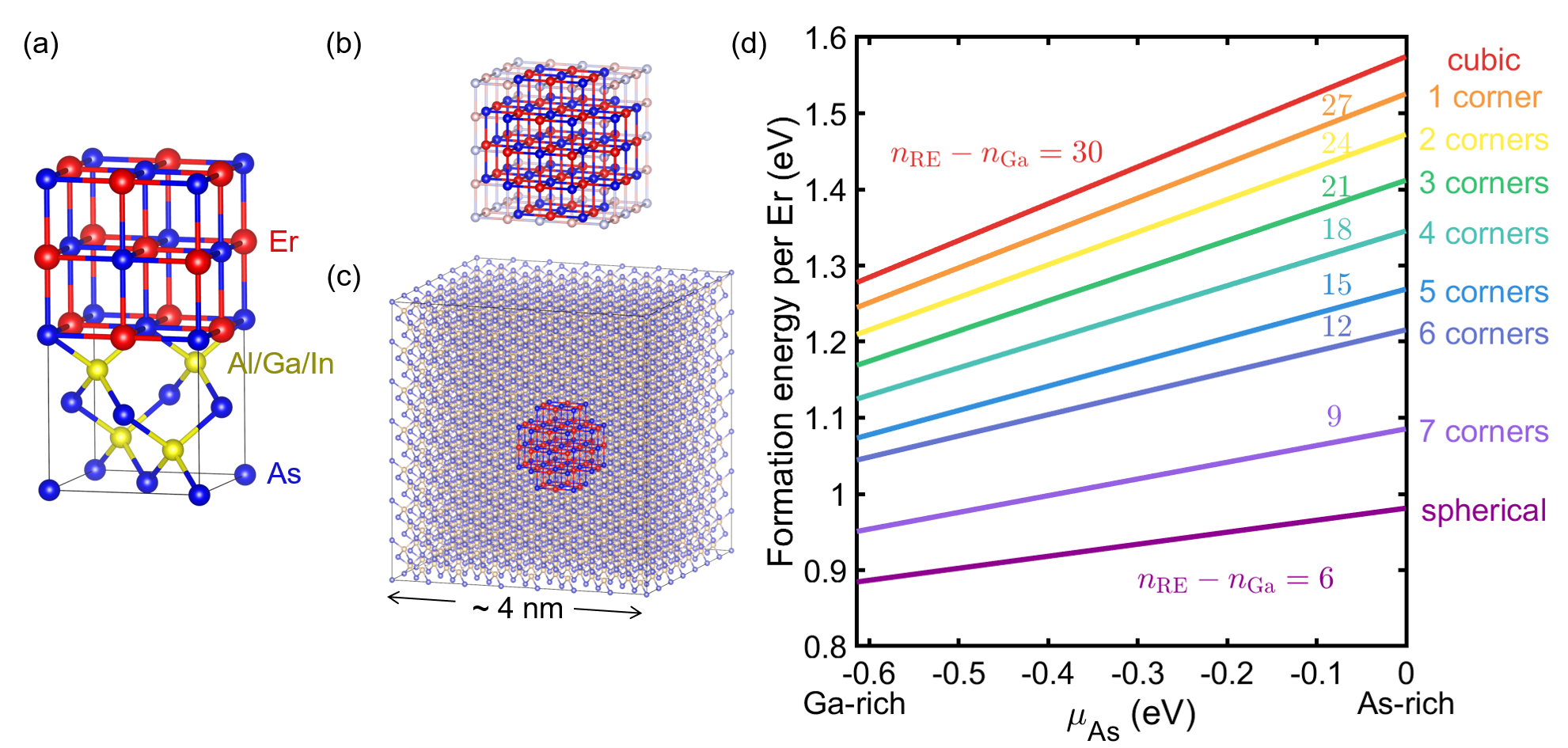}
    \captionsetup{width=\textwidth}
    \caption{Schematic representation of the (a) GaAs/ErAs(001) interface with As continuing sublattice. (b) Cubic ErAs nanoparticle with 1.14 nm side showing the corner atoms that are removed to form a spherical nanoparticle. (c) Spherical ErAs with diameter 1.14 nm in the center of a cubic supercell of the III-V matrix with approximately 4 nm edges. (d) Formation energy per Er as a function of the As chemical potential $\mu_{\rm As}$ for nine different shapes of ErAs nanoparticles in GaAs. The number of excess metal atoms ($n_{\rm RE}-n_{\rm III}$) are indicated.}
    \label{fig2}
\end{figure}

We first discuss the results for the different shapes of the embedded ErAs nanoparticles in GaAs. Transmission electron microscopy (TEM) measurements on the nanocomposite systems indicate that the embedded rare-earth monopnictide nanoparticles typically have diameters of 1-3 nm\cite{zide2005thermoelectric,zide2010high,hanson2007controlling,kawasaki2011local}. Starting from a cubic nanoparticle with a 1.14 nm side, as illustrated in Fig.~\ref{fig2}(b), we systematically generated nine distinct shapes by corner truncation, ultimately arriving at a spherical nanoparticle shape, as shown in Fig.~\ref{fig2}(c). In this procedure, we placed the center group-V atom within the RE-V nanoparticle at the group-V site at the center of the III-V host supercell, satisfying the most energetically stable GaAs/ErAs(001) interface termination (i.e., the chain model described above). 
The formation energy of the embedded nanoparticles in the III-V, $\Delta{E^f}(np)$, is defined as follows:
\begin{align}
\nonumber \Delta{E^f}(np)&=E_{tot}(np)-E_{tot}(sc)+n_{\rm III}[E_{tot}(\rm III)+\mu_{\rm III}]-\textit{n}_{\rm RE}[\textit{E}_{\textit{tot}}(\rm RE)+\mu_{\rm RE}]\\
\nonumber &=E_{tot}(np)-E_{tot}(sc)+n_{\rm III}E_{\textit{tot}}(\rm III\textnormal{-}V)-\textit{n}_{\rm RE}\textit{E}_{\textit{tot}}(\rm RE\textnormal{-}V)\\
&+(n_{\rm RE}-n_{\rm III})E_{tot}(\rm V)+(\textit{n}_{\rm RE}-\textit{n}_{\rm III})\mu_{\rm V},
\label{eq:1}
\end{align}
where $E_{tot}(np)$ is the total energy of the supercell representing the nanoparticle embedded in the semiconductor matrix, and $E_{tot}(sc)$ is the total energy of the semiconductor matrix using the same supercell; $n_{\rm III}$ is the number of group-III atoms removed from the supercell and $n_{\rm RE}$ is the number of RE atoms introduced to form the nanoparticle. Note that the number of group-V atoms remains the same, satisfying the condition of a continuous group-V sublattice as seen in the experiments~\cite{klenov2005interface}. The chemical potentials $\mu_{\rm III}$ and $\mu_{\rm RE}$ 
are variables referenced to the total energy per atom of the corresponding elemental phases and limited by the stability condition of the III-V compound and the restriction to form a continuous RE-V phase, i.e., 
\begin{align}
\mu_{\rm III}+\mu_{\rm V} = \Delta{H_f}(\rm III\textnormal{-}V)&=\textit{E}_{tot}(\rm III\textnormal{-}V)-\textit{E}_{tot}(\rm III)-\textit{E}_{tot}(\rm V),
\label{eq:2}\\
\mu_{\rm RE}+\mu_{\rm V} < \Delta{H_f}(\rm RE\textnormal{-}V)&=\textit{E}_{tot}(\rm RE\textnormal{-}V)-\textit{E}_{tot}(\rm RE)-\textit{E}_{tot}(\rm V),
\label{eq:3}
\end{align}
where $\Delta{H_f}(\rm III\textnormal{-}V)$ is the formation enthalpy of the III-V semiconductor and $\Delta{H_f}(\rm RE\textnormal{-}V)$ is the formation enthalpy of the RE-V semimetal. The formation energy $\Delta{E^f}(np)$ per Er as a function of the As chemical potential $\mu_{\rm As}$ for the nine different shapes of the ErAs nanoparticles in GaAs are shown in Fig~.\ref{fig2}(d). We find that the formation energy decreases monotonically with the number of corners removed, from cubic to spherical, and that a sphere is the lowest energy shape of the nanoparticle. In other words, spherical ErAs nanoparticles are more likely to form than cubic-shaped ones, which explains the reported observations based on TEM measurements \cite{zide2005thermoelectric,zide2010high}. We also note that the spherical nanoparticle also leads to the lowest density of the nanocomposite material, minimizing the number of excess metal atoms ($n_{\rm RE}-n_{\rm III}$).

\begin{figure}
\centering
\includegraphics[width=\textwidth]{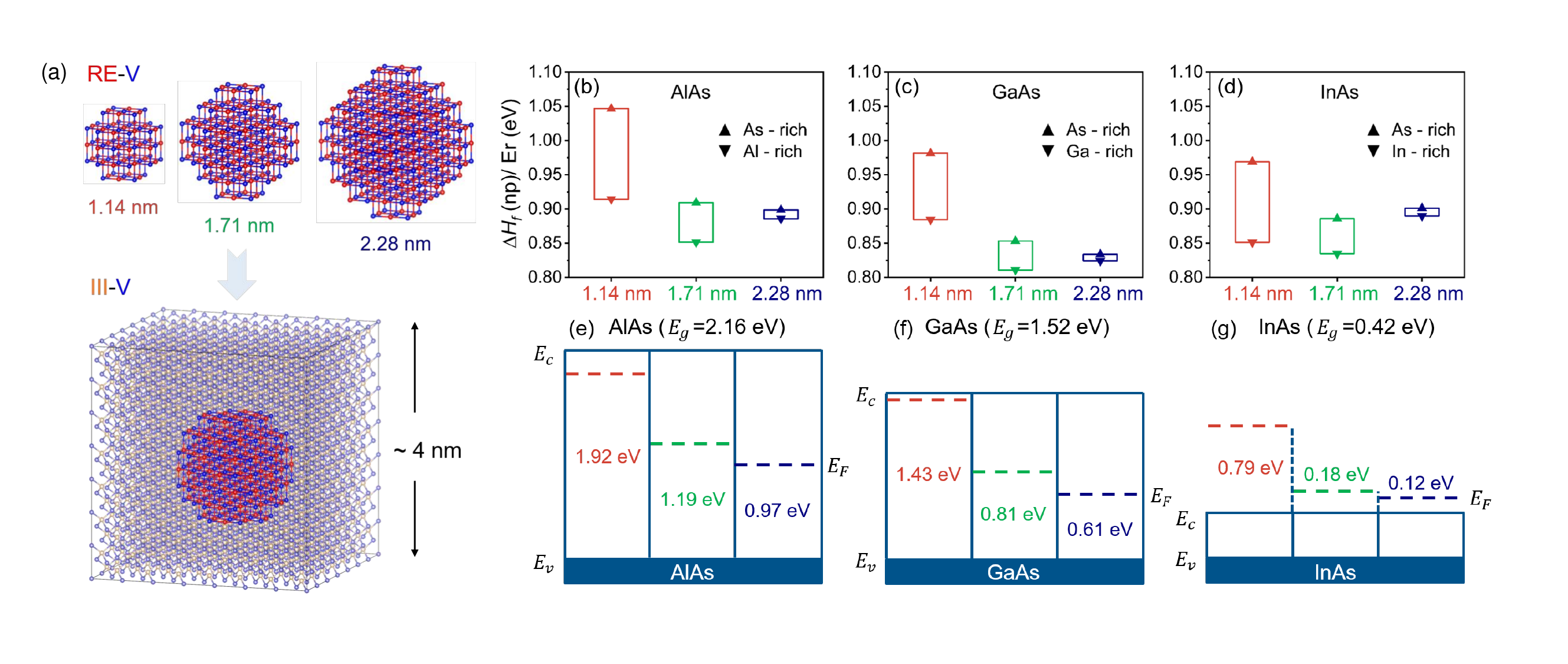}
\captionsetup{width=\textwidth}
\caption{\label{fig3}
ErAs nanoparticle sizes in III-Vs and Fermi-level pinning. (a) Ball \& stick model of three different sizes, 1.14 nm, 1.71 nm and 2.28 nm, of ErAs spherical nanoparticles in the approximately 4 nm III-V cubic supercell matrix. Formation energy per Er for the three sizes of ErAs nanoparticles in three III-V matrices: (b) AlAs, (c) GaAs and (d) InAs. Here, $\blacktriangle$ represents As-rich condition, i.e., $\mu_{\rm As}=0$, and $\blacktriangledown$ corresponds Al-rich, Ga-rich or In-rich condition, representing $\mu_{\rm Al}=0$ ($\mu_{\rm As}=\Delta{H_f}(\rm AlAs)$), $\mu_{\rm Ga}=0$ ($\mu_{\rm As}=\Delta{H_f}(\rm GaAs)$) and $\mu_{\rm In}=0$ ($\mu_{\rm As}=\Delta{H_f}(\rm InAs)$). Fermi level ($E_F$) in the composite materials with respect to the valence-band maximum $E_v$ in (e) AlAs and (f) GaAs, and conduction-band minimum $E_c$ in (g) InAs.}
\end{figure}

Taking spherical as the most favorable shape of the nanoparticles, we then explore how the formation energy varies with size, considering three nanopaticles sizes, of 1.14 nm, 1.71 nm, and 2.28 nm in diameter, in the approximately 4 nm cubic supercell of the III-V semiconductor matrix. These configurations are displayed in Fig.~\ref{fig3}(a), and the calculated formation energies are shown in Fig.~\ref{fig3}(b), (c) and (d) for ErAs nanoparticles in AlAs, GaAs, and InAs, respectively. In these diagrams, the height of the rectangles corresponds to the range of the chemical potential $\mu_{\rm As}$, varying from As-poor (or III-rich, bottom) to As-rich (III-poor, top) limiting conditions. The results show that the 1.71 nm and 2.28 nm nanoparticles have lower formation energy than the 1.14 nm nanoparticle and that a minimum occurs for nanoparticles with sizes between 1.71 and 2.28 nm in diameter taking the middle point between the As-rich and As-poor limit conditions.
Note, first, that in the limit of very large particle sizes (diameter $\rightarrow \infty$) the formation energy should converge to a number related to the interface energy per Er. Second, these results indicate that the most energetically favorable particle size lies between 1.71 nm and 2.28 nm, and that larger nanoparticle sizes would be less energetically favorable. It is worth noting that we do not rule out that larger embedded nanoparticles could not be fabricated, as that would depend on the amount of Er introduced into GaAs. Their formation could be attributed to kinetic mechanisms. Our results indicate that thermodynamically, 1.71 nm and 2.28 nm nanoparticles would be more energetically favorable than smaller or larger nanoparticles. This effect is most likely attributed to the relationship between interface energy and the bulk energy of the RE-V semimetal, and further studies would be required. Lastly, experimental observations point to the nanoparticles in III-Vs having diameters approximating 2nm \cite{hanson2007controlling,kim2008reducing,liu2011properties,kawasaki2011local}.
  
Finally, the electronic structure of the embedded RE-V nanoparticles in the III-V matrices was determined by calculating the atom-resolved density of states (DOS) of the composite materials. The Fermi level in the nanocomposite material was then aligned to the band edges of the III-V parent compounds. Here, we assume that the As located farthest away from the center of the nanoparticle in the supercell representing the composite material is a bulk-like As in the semiconductor matrix. The alignment is then performed by carrying out two calculations: In the first calculation, we obtain the Fermi level in the nanocomposite material with respect to the electrostatic potential averaged over a sphere (2.12 \AA~in diameter) around the As atom farthest away from the nanoparticle. In the second calculation, we obtain the valence-band-maximum $E_v$ with respect to the electrostatic potential averaged over a sphere around one As atom in the semiconductor material, using either the supercell representing the pristine material or the primitive cell (the results are the same). From these two calculations, we determine the Fermi level of the embedded nanoparticle composite material with respect to $E_v$ in the semiconductor. The results for ErAs embedded in GaAs are shown in Fig.~\ref{fig3}(e). 

Based on this alignment, the Fermi level ($E_F$) moves towards $E_v$ of the III-V as the size of the ErAs nanoparticle increases. For the smallest nanoparticle (1.14 nm), $E_F$ is close to the conduction-band minimum $E_c$, whereas, for the larger sizes of 1.71 nm and 2.28 nm, $E_F$ is located deep within the band gap of the III-V. In particular, for the 1.71 nm nanoparticles, $E_F$ is pinned at mid-gap of GaAs, consistent with previous experiments based on X-ray scanning tunneling microscopy/spectroscopy measurements (XSTM/XSTS) \cite{kawasaki2011local}. 
Note that the 1.71 nm nanoparticles are also closer in size to those observed in the TEM measurements \cite{zide2005thermoelectric,zide2010high}. The resulting deep states lead to short carrier lifetimes and high dark resistance, desirable properties for large bandwidth and a high signal-to-noise ratio, indicating that these nanoparticle systems are promising candidates for THz source/detector materials \cite{bomberger2017overview,o2006enhanced}.

\begin{figure}
\centering
\includegraphics[width=\textwidth]{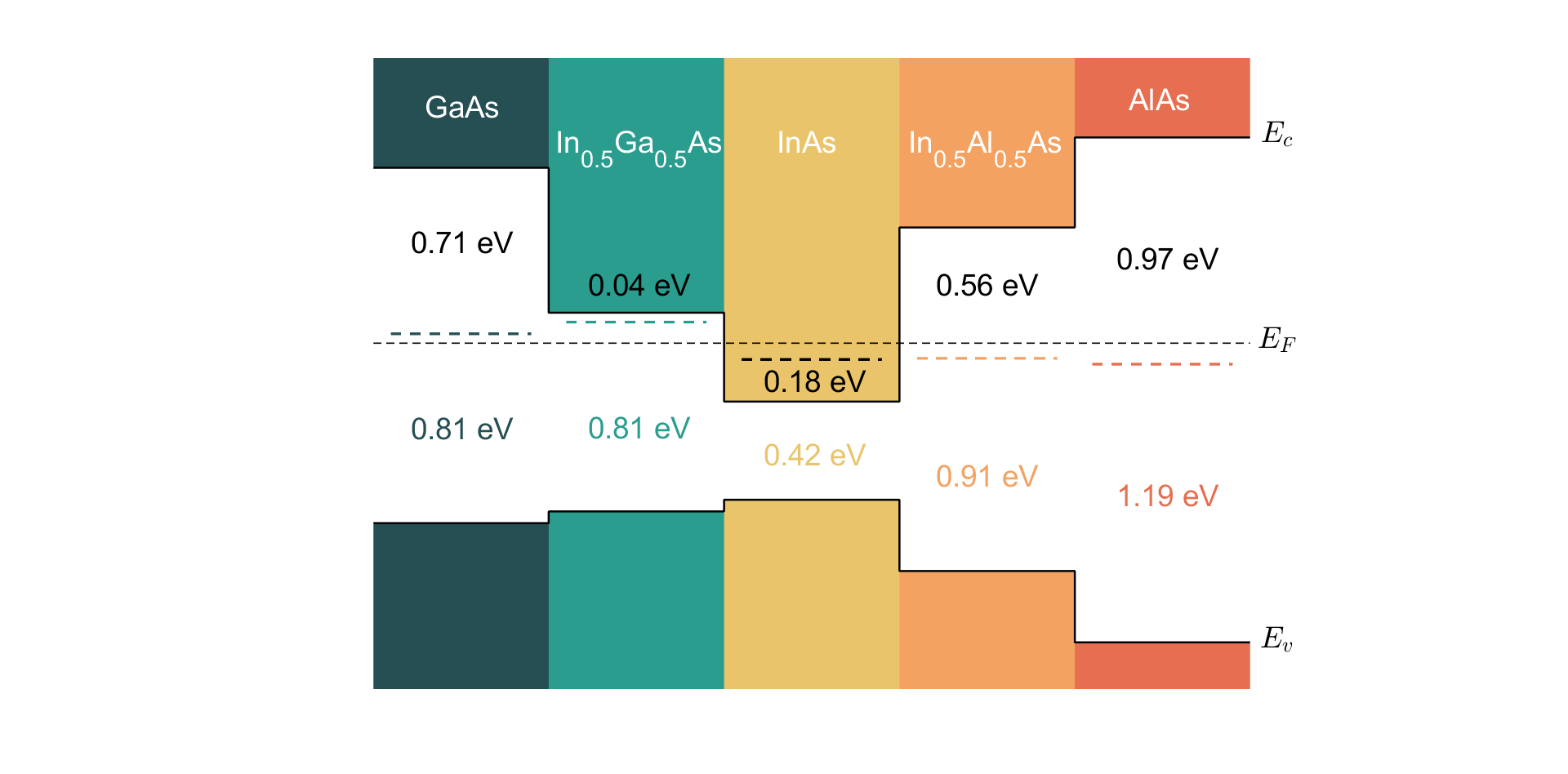}
\captionsetup{width=\textwidth}
\caption{\label{fig4}
Fermi level ($E_F$) pinning for 1.71 nm embedded ErAs nanoparticles in different III-Vs and their alloys. The band alignment of III-Vs AlAs, GaAs, and InAs are based on HSE06 calculations from the literature \cite{gruneis2014ionization,harrington2017valence}.}
\end{figure}

We note that the results in Fig.~\ref{fig3}(e)-(g) display an interesting trend. If the known band offsets between the III-V semiconductors AlAs, GaAs, and InAs are taken into account\cite{gruneis2014ionization,harrington2017valence}, the Fermi level in the nanocomposite material, e.g. with 1.71 nm nanoparticles in the different III-V semiconductors are pinned at the same energy with respect to the vacuum level, as shown in Fig.~\ref{fig4}. This result can also be extended to the alloys of these three semiconductor materials, also included in Fig.~\ref{fig4}. In these calculations, the ternary alloys In$_{0.5}$Al$_{0.5}$As and In$_{0.5}$Ga$_{0.5}$As are modeled using special quasi-random structures (SQSs) \cite{zunger1990special}, with In and Ga (Al) atoms distributed over the group-III lattice sites to represent a random alloy using the 2744-atom cubic supercell. In our calculations, we chose In$_{0.5}$Al$_{0.5}$As and In$_{0.5}$Ga$_{0.5}$As based on the fact that In$_{0.52}$Al$_{0.48}$As and In$_{0.53}$Ga$_{0.47}$As have been used in experimental growth because they are lattice-matched with InP substrates\cite{zide2005thermoelectric,klenov2005interface,kim2006thermal}. 

\begin{figure}
\centering
\includegraphics[width=0.9\textwidth]{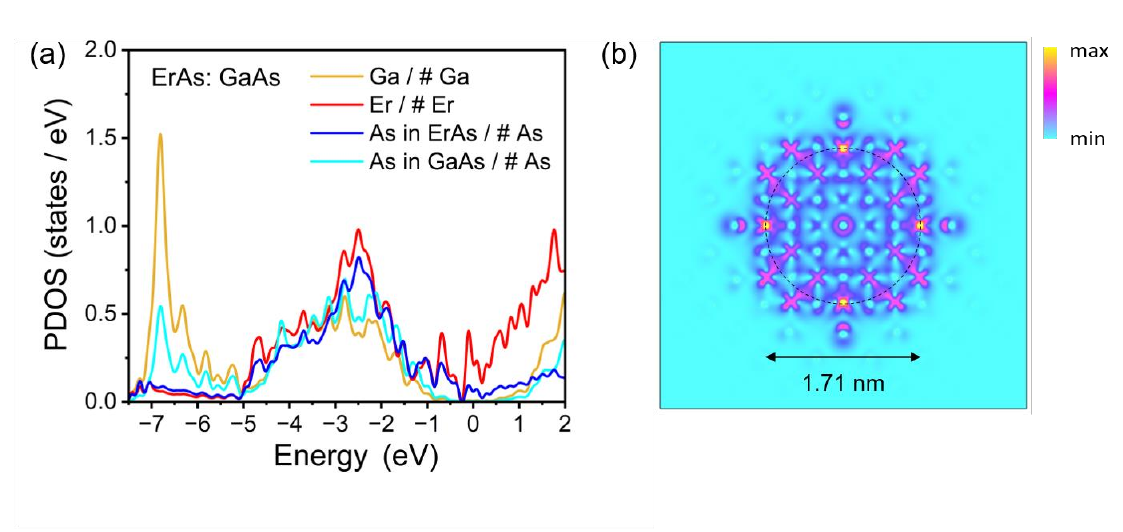}
\captionsetup{width=\textwidth}
\caption{\label{fig5}
Projected density of states (PDOS) of a supercell of GaAs containing an ErAs nanoparticle of 1.71 nm diameter. (a) The average of the PDOS is calculated for all Ga atoms in the matrix and all Er atoms in the nanoparticle. PDOS of the As atoms is separated into two parts, corresponding to either in the GaAs matrix or the ErAs nanoparticle. The Fermi level is set to zero. (b) Charge density isosurface of the states around $E_F$ ($\pm$ 0.1 eV) in the (001) plane crossing the center of the ErAs nanoparticle. Similar plots for other matrices can be found in the SI.}
\end{figure}

The pinning of the Fermi level $E_F$ in an absolute energy scale for the ErAs-based nanocomposite materials can be understood by analyzing the atom-resolved density of states near $E_F$. Taking the case of GaAs as an example, we show in Fig.~\ref{fig5} the calculated density of states for the GaAs region representing the bulk-like properties and the density of states projected on the atoms that compose the ErAs particle. The Fermi level, in the middle of the gap of the GaAs matrix, crosses ErAs-related states, as shown in Fig.~\ref{fig5}(a), with the largest contribution from the Er atoms. Focusing on the electronic states in the vicinity of the $E_F$, within a range of approximately $\pm$0.1 eV, the charge density associated with these states reveals predominant contribution from the ErAs nanoparticle, particularly its outermost layer (Fig.~\ref{fig5}(b)).

Based on the results above, we conclude that the Fermi level pinning is related solely to the RE-V that compose the nanoparticle, and that occurs near the mid gap in GaAs. Changing the III-V semiconductor will only change the band edges of the host material, offering a design principle for tuning the Fermi-level pinning in the composite material for specific applications. These results can also be generalized to other RE-V compounds (except Ce and Eu containing compounds due to the f states crossing the Fermi level, leading to more complex behavior). The specific position of the Fermi level, for a given nanoparticle size in different III-V matrices, could be tuned by changing the RE-V material, choosing from LaAs to LuAs, independent of the III-V matrices. Changing the III-V matrices will change the band-edge positions of the host material and, of course, the overall properties of the composite material.

\section{Summary}

We use first-principles calculations based on DFT and the meta-GGA mBJ functional to investigate the electronic structure of semimetal ErAs nanoparticles with different shapes and sizes embedded in III-V (AlAs, GaAs and InAs) and their ternary alloys. Our results show that, for nanoparticles of the same size, the spherical shape is the most energetically favorable. As the size of the ErAs nanoparticle increases, from 1.14 nm to 1.71 nm and 2.28 nm, the Fermi level shifts towards the valence band of the III-V. Specifically, the 1.71 nm spherical ErAs nanoparticle pins the Fermi level near the mid-gap in GaAs, and provides an explanation to previous experimental observations. Moreover, the Fermi level is pinned at the same energy in an absolute energy scale, considering the band offsets between the III-Vs and their alloys. Our calculations provide valuable insights for the selection of RE-V and III-V matrices in the nanoparticle embedded composite materials in designing novel composite materials with target properties for device applications.

\section{Acknowledgements}
This work was supported by the NSF through the UD-CHARM University of Delaware Materials Research Science and Engineering Center (MRSEC) grant No. DMR-2011824. We also acknowledge the use of Bridges-2 at PSC through allocation DMR150099 from the Advanced Cyberinfrastructure Coordination Ecosystem: Services \& Support (ACCESS) program, which is supported by National Science Foundation grants No. 2138259, 2138286, 2138307, 2137603, and 2138296, and the DARWIN computing system at the University of Delaware, which is supported by the NSF grant No.~1919839.

\bibliography{ErAs-NP}
\end{document}


\section{Note S1. Density of states and charge density of ErAs nanoparticle embedded in different III-Vs and their alloys}

Due to the large number of atoms for the nanoparticle composite materials, only the $\Gamma$ point is used when computing the density of states of the composite systems.

\begin{figure} 
    \centering
    \includegraphics[width=0.87\textwidth]{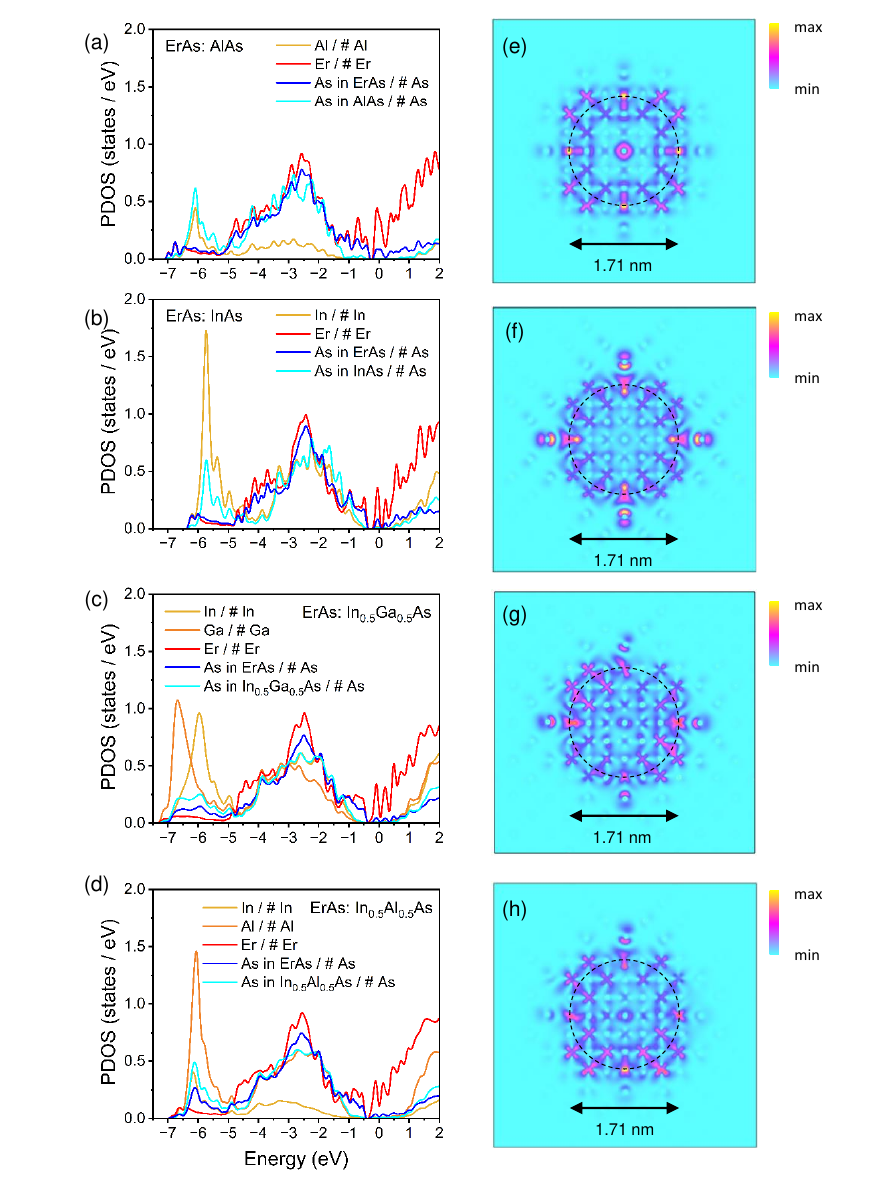}
    \captionsetup{width=\textwidth}
    \caption{Projected density of states (PDOS) of the 1.71 nm ErAs nanoparticle embedded in around 4 nm size (a) AlAs, (b) InAs, and III-V alloys (c) \ch{In$_{0.5}$Ga$_{0.5}$As} and (d) \ch{In$_{0.5}$Al$_{0.5}$As} matrices. Here, PDOS of the As atoms is separated into two parts, corresponding to either in the III-V matrix or the ErAs nanoparticle. PDOS of the Group III elements or Er atoms represents the average value obtained by dividing by the number of atoms of that element. The Fermi level is set to zero. (e)-(h) Charge density isosurface of the states around $E_F$ ($\pm$ 0.1 eV) in the (001) plane crossing the center of the ErAs nanocomposite with AlAs, InAs, \ch{In$_{0.5}$Ga$_{0.5}$As} and \ch{In$_{0.5}$Al$_{0.5}$As} matrices respectively.}
    \label{figS1}
\end{figure}